\documentclass[trackchanges]{aastex7} 
\suppressAffiliationstrue
\usepackage{graphicx}

\usepackage{colortbl}  
\definecolor{lightgrey}{gray}{0.95}
\usepackage[table]{xcolor}
\begin{document}

\title{Enhancing Multiplet Alignment Measurements with Imaging}




\author[orcid=0009-0009-8150-684X,sname='North America']{Alexus Annika Kumwembe}
\affiliation{Center for Astrophysics $|$ Harvard \& Smithsonian}
\affiliation{Saint Mary's College of Maryland}
\email[show]{aakumwembe@gmail.com} 

\author[orcid=0000-0002-6731-9329]{Claire Lamman}
\affiliation{Center for Astrophysics $|$ Harvard \& Smithsonian, 60 Garden Street, Cambridge, MA 02138, USA}
\affiliation{The Ohio State University, Columbus, 43210 OH, USA}
\email{lamman.1@osu.edu}

\author{Daniel Eisenstein}
\affiliation{Center for Astrophysics $|$ Harvard \& Smithsonian, 60 Garden Street, Cambridge, MA 02138, USA}
\email{deisenstein@cfa.harvard.edu}

\author{Jessica Nicole Aguilar}
\affiliation{Lawrence Berkeley National Laboratory, 1 Cyclotron Road, Berkeley, CA 94720, USA}
\email{jaguilar@lbl.gov}

\author[orcid=0000-0001-6098-7247]{Steven Ahlen}
\affiliation{Department of Physics, Boston University, 590 Commonwealth Avenue, Boston, MA 02215 USA}
\email{ahlen@bu.edu}

\author[orcid=0000-0001-9712-0006]{Davide Bianchi}
\affiliation{Dipartimento di Fisica ``Aldo Pontremoli'', Universit\`a degli Studi di Milano, Via Celoria 16, I-20133 Milano, Italy}
\affiliation{INAF-Osservatorio Astronomico di Brera, Via Brera 28, 20122 Milano, Italy}
\email{davide.bianchi1@unimi.it}

\author{David Brooks}
\affiliation{Department of Physics \& Astronomy, University College London, Gower Street, London, WC1E 6BT, UK}
\email{david.brooks@ucl.ac.uk}

\author{Todd Claybaugh}
\affiliation{Lawrence Berkeley National Laboratory, 1 Cyclotron Road, Berkeley, CA 94720, USA}
\email{tmclaybaugh@lbl.gov}

\author[orcid=0000-0002-2169-0595]{Andrei Cuceu}
\affiliation{Lawrence Berkeley National Laboratory, 1 Cyclotron Road, Berkeley, CA 94720, USA}
\email{acuceu@lbl.gov}

\author[orcid=0000-0002-1769-1640]{Axel de la Macorra}
\affiliation{Instituto de F\'{i}sica, Universidad Nacional Aut\'{o}noma de M\'{e}xico,  Circuito de la Investigaci\'{o}n Cient\'{i}fica, Ciudad Universitaria, Cd. de M\'{e}xico  C.~P.~04510,  M\'{e}xico}
\email{macorra@fisica.unam.mx}

\author[orcid=0000-0002-5665-7912]{Biprateep Dey}
\affiliation{Department of Astronomy \& Astrophysics, University of Toronto, Toronto, ON M5S 3H4, Canada}
\affiliation{Department of Physics \& Astronomy and Pittsburgh Particle Physics, Astrophysics, and Cosmology Center (PITT PACC), University of Pittsburgh, 3941 O'Hara Street, Pittsburgh, PA 15260, USA}
\email{b.dey@utoronto.ca}

\author{Peter Doel}
\affiliation{Department of Physics \& Astronomy, University College London, Gower Street, London, WC1E 6BT, UK}
\email{apd@star.ucl.ac.uk}

\author[orcid=0000-0002-3033-7312]{Andreu Font-Ribera}
\affiliation{Institut de F\'{i}sica d'Altes Energies (IFAE), The Barcelona Institute of Science and Technology, Edifici Cn, Campus UAB, 08193, Bellaterra (Barcelona), Spain}
\email{afont@ifae.es}

\author[orcid=0000-0002-2890-3725]{Jaime E. Forero-Romero}
\affiliation{Departamento de F\'isica, Universidad de los Andes, Cra. 1 No. 18A-10, Edificio Ip, CP 111711, Bogot\'a, Colombia}
\affiliation{Observatorio Astron\'omico, Universidad de los Andes, Cra. 1 No. 18A-10, Edificio H, CP 111711 Bogot\'a, Colombia}
\email{je.forero@uniandes.edu.co}

\author[orcid=0000-0001-9632-0815]{Enrique Gaztañaga}
\affiliation{Institut d'Estudis Espacials de Catalunya (IEEC), c/ Esteve Terradas 1, Edifici RDIT, Campus PMT-UPC, 08860 Castelldefels, Spain}
\affiliation{Institute of Cosmology and Gravitation, University of Portsmouth, Dennis Sciama Building, Portsmouth, PO1 3FX, UK}
\affiliation{Institute of Space Sciences, ICE-CSIC, Campus UAB, Carrer de Can Magrans s/n, 08913 Bellaterra, Barcelona, Spain}
\email{gaztanaga@gmail.com}

\author[orcid=0000-0003-3142-233X]{Satya Gontcho A Gontcho}
\affiliation{Lawrence Berkeley National Laboratory, 1 Cyclotron Road, Berkeley, CA 94720, USA}
\affiliation{University of Virginia, Department of Astronomy, Charlottesville, VA 22904, USA}
\email{satya@virginia.edu}

\author{Gaston Gutierrez}
\affiliation{Fermi National Accelerator Laboratory, PO Box 500, Batavia, IL 60510, USA}
\email{gaston@fnal.gov}

\author[orcid=0000-0002-6024-466X]{Mustapha Ishak}
\affiliation{Department of Physics, The University of Texas at Dallas, 800 W. Campbell Rd., Richardson, TX 75080, USA}
\email{mishak@utdallas.edu}

\author[orcid=0000-0001-8528-3473]{Jorge Jimenez}
\affiliation{Institut de F\'{i}sica d'Altes Energies (IFAE), The Barcelona Institute of Science and Technology, Edifici Cn, Campus UAB, 08193, Bellaterra (Barcelona), Spain}
\email{jjimenez@ifae.es}

\author[orcid=0000-0003-0201-5241]{Dick Joyce}
\affiliation{NSF NOIRLab, 950 N. Cherry Ave., Tucson, AZ 85719, USA}
\email{richard.joyce@noirlab.edu}

\author{Robert Kehoe}
\affiliation{Department of Physics, Southern Methodist University, 3215 Daniel Avenue, Dallas, TX 75275, USA}
\email{kehoe@physics.smu.edu}

\author[orcid=0000-0003-3510-7134]{Theodore Kisner}
\affiliation{Lawrence Berkeley National Laboratory, 1 Cyclotron Road, Berkeley, CA 94720, USA}
\email{tskisner@lbl.gov}

\author{Ofer Lahav}
\affiliation{Department of Physics \& Astronomy, University College London, Gower Street, London, WC1E 6BT, UK}
\email{o.lahav@ucl.ac.uk}

\author[orcid=0000-0003-1838-8528]{Martin Landriau}
\affiliation{Lawrence Berkeley National Laboratory, 1 Cyclotron Road, Berkeley, CA 94720, USA}
\email{mlandriau@lbl.gov}

\author[orcid=0000-0003-4962-8934]{Marc Manera}
\affiliation{Departament de F\'{i}sica, Serra H\'{u}nter, Universitat Aut\`{o}noma de Barcelona, 08193 Bellaterra (Barcelona), Spain}
\affiliation{Institut de F\'{i}sica d'Altes Energies (IFAE), The Barcelona Institute of Science and Technology, Edifici Cn, Campus UAB, 08193, Bellaterra (Barcelona), Spain}
\email{mmanera@ifae.es}

\author{Ramon Miquel}
\affiliation{Instituci\'{o} Catalana de Recerca i Estudis Avan\c{c}ats, Passeig de Llu\'{i}s Companys, 23, 08010 Barcelona, Spain}
\affiliation{Institut de F\'{i}sica d'Altes Energies (IFAE), The Barcelona Institute of Science and Technology, Edifici Cn, Campus UAB, 08193, Bellaterra (Barcelona), Spain}
\email{rmiquel@ifae.es}

\author[orcid=0000-0001-9070-3102]{Seshadri Nadathur}
\affiliation{Institute of Cosmology and Gravitation, University of Portsmouth, Dennis Sciama Building, Portsmouth, PO1 3FX, UK}
\email{seshadri.nadathur@port.ac.uk}

\author[orcid=0000-0003-3188-784X]{Nathalie Palanque-Delabrouille}
\affiliation{IRFU, CEA, Universit\'{e} Paris-Saclay, F-91191 Gif-sur-Yvette, France}
\affiliation{Lawrence Berkeley National Laboratory, 1 Cyclotron Road, Berkeley, CA 94720, USA}
\email{npalanque-delabrouille@lbl.gov}

\author[orcid=0000-0001-6979-0125]{Ignasi Pérez-Ràfols}
\affiliation{Departament de F\'isica, EEBE, Universitat Polit\`ecnica de Catalunya, c/Eduard Maristany 10, 08930 Barcelona, Spain}
\email{ignasi.perez.rafols@upc.edu}

\author[orcid=0000-0001-7145-8674]{Francisco Prada}
\affiliation{Instituto de Astrof\'{i}sica de Andaluc\'{i}a (CSIC), Glorieta de la Astronom\'{i}a, s/n, E-18008 Granada, Spain}
\email{fprada@iaa.es}

\author{Graziano Rossi}
\affiliation{Department of Physics and Astronomy, Sejong University, 209 Neungdong-ro, Gwangjin-gu, Seoul 05006, Republic of Korea}
\email{graziano@sejong.ac.kr}

\author[orcid=0000-0002-9646-8198]{Eusebio Sanchez}
\affiliation{CIEMAT, Avenida Complutense 40, E-28040 Madrid, Spain}
\email{eusebio.sanchez@ciemat.es}

\author{David Schlegel}
\affiliation{Lawrence Berkeley National Laboratory, 1 Cyclotron Road, Berkeley, CA 94720, USA}
\email{djschlegel@lbl.gov}

\author[orcid=0000-0002-6588-3508]{Hee-Jong Seo}
\affiliation{Department of Physics \& Astronomy, Ohio University, 139 University Terrace, Athens, OH 45701, USA}
\email{seoh@ohio.edu}

\author[orcid=0000-0002-3461-0320]{Joseph Harry Silber}
\affiliation{Lawrence Berkeley National Laboratory, 1 Cyclotron Road, Berkeley, CA 94720, USA}
\email{jhsilber@lbl.gov}

\author{David Sprayberry}
\affiliation{NSF NOIRLab, 950 N. Cherry Ave., Tucson, AZ 85719, USA}
\email{david.sprayberry@noirlab.edu}

\author[orcid=0000-0003-1704-0781]{Gregory Tarlé}
\affiliation{University of Michigan, 500 S. State Street, Ann Arbor, MI 48109, USA}
\email{gtarle@umich.edu}

\author{Benjamin Alan Weaver}
\affiliation{NSF NOIRLab, 950 N. Cherry Ave., Tucson, AZ 85719, USA}
\email{benjamin.weaver@noirlab.edu}

\begin{abstract}

We demonstrate that measurements of the gravitational tidal field made with spectroscopic redshifts can be improved with information from imaging surveys. The average orientation of small groups of galaxies, or “multiplets” is correlated with large-scale structure and is used to measure the direction of tidal forces. Previously, multiplet intrinsic alignment has been measured in DESI using galaxies that have spectroscopic redshifts. The DESI Legacy Imaging catalog can be used to supplement multiplet catalogs. Our findings show that galaxy positions from the imaging catalog produce a measurement similar to the measurements made with only spectroscopic data. This demonstrates that imaging can improve our signal-to-noise ratio for multiplet alignment in DESI.

\end{abstract}



\section{Introduction} 

As large-scale structure grows, the tidal fields that form these structures induce effects on the shape, spin, and orientation of galaxies. These effects are classified as Intrinsic Alignments (IA) \citep{joachimi_intrinsic_2013, troxel_IA_2015}. IA traces cosmological effects on the large-scale tidal field \citep{chisari_cosmological_2013}, including correlating shapes 
to the quadrupole of projected large-scale structure.  
IA also applies to galaxy ensembles, such as galaxy groups, clusters, and halos \citep{rong_2025, shi_intrinsic_2024, schneider_shapes_2012}. 

To maximize the number of galaxy ensembles, multiplets can be used. These are smaller than groups and not necessarily virialized.
Multiplets can be better for measuring alignment than individual galaxies when samples: have poor imaging, are dense, or have little individual alignment \citep{lamman_detection_2024}. Multiplet alignment is a projected quantity, correlating multiplets' orientations in the plane of the sky with the projected separation to tracer galaxies. Multiplets can be constructed without the full 3D positions of each multiplet member, since multiplet orientation is obtained from projected positions. 
Here, we explore whether multiplets made with imaging and spectroscopic data can produce the same signal as multiplets identified with only spectroscopic data.

\section{Data} \label{sec:style}


DESI constrains structure growth and dark energy by obtaining millions of 
extragalactic redshifts \citep{DESI2024.VI.KP7A, DESI2024.IV.KP6, DESI2024.III.KP4, DESI2024.II.KP3}. We use two catalogs: the DESI Y1 Luminous Red Galaxy (LRG) Sample \citep{zhou_target_2023}, which contains spectroscopic redshifts, and the DESI Legacy Imaging Survey \citep{dey_overview_2019, desi_collaboration_overview_2022} which consists of the imaging information used in DESI's target selection, including $r$ and $W1$ photometry \citep{DESI2016a.Science}. 

We chose LRGs since they have high bias and display strong multiplet alignment.
DESI's Y1 catalog consists of 8 million LRGs in the redshift range of 0.4 $< z < \sim$ 1.0. It is the same survey used in the initial detection of multiplet alignment \citep{lamman_detection_2024}, allowing us to make a comparison to the spectroscopic only case. 
From the imaging catalog, we select LRG-like galaxies that are twice as faint as the galaxies in the spectroscopic catalog
 by using original cuts for the main survey, but with a lower fiber magnitude cut of $z<22.35$. Our resulting
LRG imaging sample has a density 9.4 times higher than the spectroscopic sample (605 deg$^{-2}$). 

\section{Methods} \label{sec:methods}

We identify imaged galaxies close to spectroscopic targets and measure 
the alignment of these multiplets relative to positions of the full LRG spectroscopic sample.
 We explore three limits on the sky distance between imaged and spectroscopic galaxies: 0.6 arcminutes, 1.8 arcminutes, and 3 arcminutes. These correspond to a transverse comoving distance of about 0.3 - 1 Mpc in the LRG redshift range, the distance used to define spectroscopic multiplets.

 The members of multiplets are further limited by the difference in color compared to the spectroscopic galaxy, increasing the probability that they are at the same redshift. 
 The two color-based cuts we explore are $\Delta (r-W1)$ less than 0.1 and 0.3. From combinations of these cuts, we create six enhanced multiplet catalogs. 

The multiplet alignment signal is defined and calculated following \cite{lamman_detection_2024} which we summarize here. The key difference is that we assign the imaged neighboring galaxies the same redshift as their center galaxy. The projected orientation of multiplets is determined from members' sky positions, then correlated with the positions of all LRGs in the spectroscopic catalog. The distance along the line of sight between the multiplet’s center and a tracer, $\Pi_{\rm max}$, increases with their projected separation, $R$.
Alignment is described by $\cos(2\phi)$, measured as a function of $R$. $\phi$ is the projected multiplet orientation relative to tracers.

\begin{figure} 
\centering
\includegraphics[width=\textwidth]{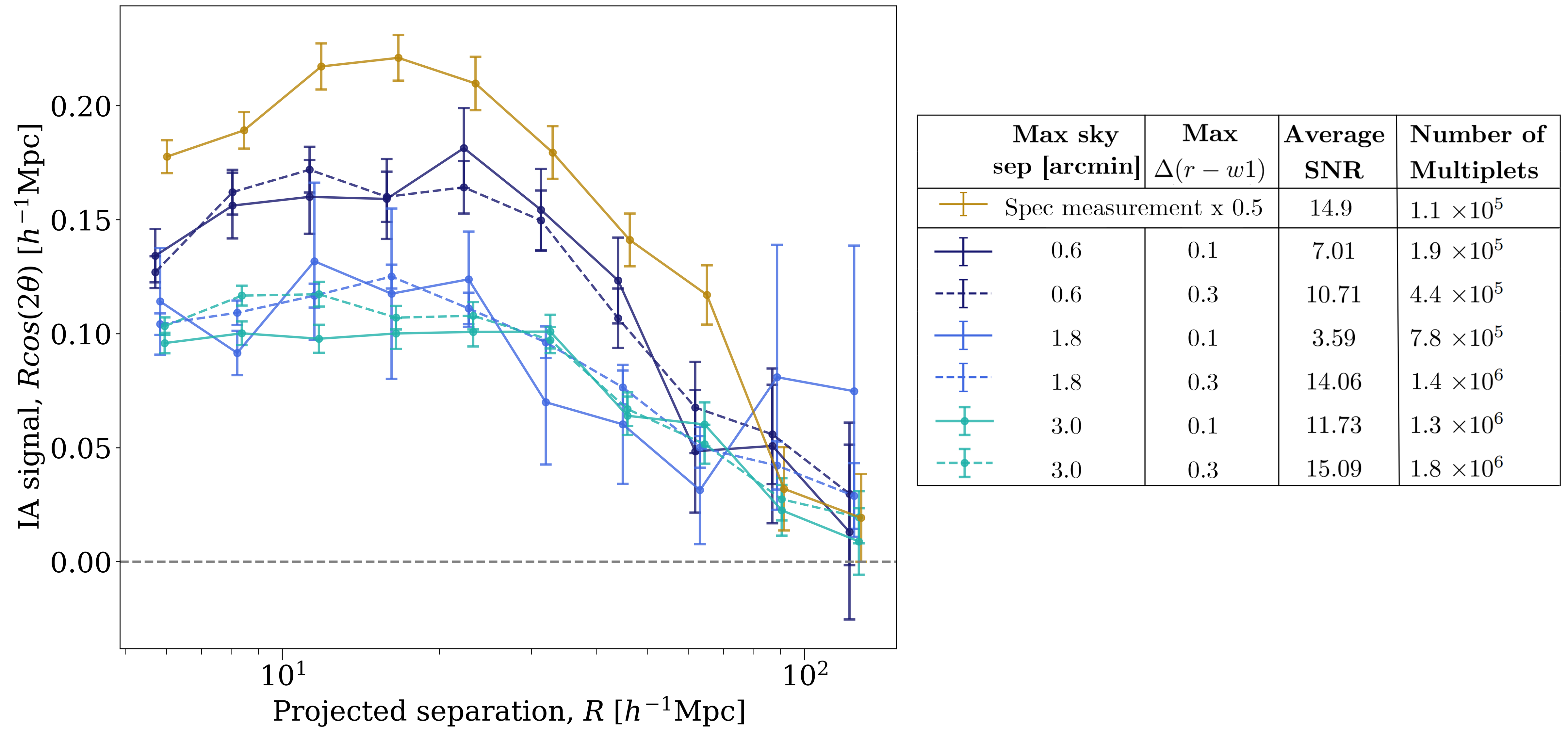}  
\caption{
The tidal alignment of galaxy multiplets, each containing one spectroscopically-measured central galaxy.
The gold line is the original measurement made with only spectroscopic data. 
Although amplitudes vary depending on multiplet definition, their scale-dependent response to large-scale shear is consistent.
} 
\label{fig:main_figure} 
\end{figure}

\section{Results}

Our results demonstrate that supplementing a redshift catalog with imaging to find multiplets can produce a signal on par with spectroscopic-only measurements.
Figure \ref{fig:main_figure} compares the original multiplet alignment from the spectroscopic catalog to 
measurements of the enhanced catalogs. The signals have the same scale dependence 
with comparable signal-to-noise ratios (SNR).
Multiplets made with the strictest cuts 
produce the most similar signal to the spectroscopic case. 
This `purest' multiplet catalog produces the highest correlations, but a more generous cut will produce more multiplets and a higher SNR. Notably, 
the least strict cuts produce a higher SNR than the spectroscopic measurement. 

The main factors affecting the signal amplitude are the number of unassociated photometric interlopers and the multiplet size. 
The alignment of multiplets made with larger cuts are more likely to be diluted by interlopers, but are also larger, which increases alignment strength.
While the alignment amplitude varies between our samples, they display identical scale dependence in the linear regime. This can be directly related to the underlying tidal shear for some applications and the amplitude calibrated or otherwise inferred for others. For a more detailed discussion of applications, see \cite{lamman_detection_2024}.


This work presents a proof-of-concept and initial suggestions for which cuts will optimize future work combining imaging with spectroscopic data. 
Our technique can be used for galaxy populations besides LRGs, and 
for denser imaging surveys such as LSST \citep{ivezic_lsst_2019}. This has
the potential to outperform multiplet measurements made with spectra alone.
\\

\noindent \textit{Data plotted here can be found at \href{https://zenodo.org/uploads/17298578}{zenodo.org/uploads/17298578}.}



\section{Acknowledgments}
This material is based upon work supported by the U.S.\ Department of Energy under grants DE-SC0013718 and DE-SC0024787, and the Research Experiences for Undergraduates Program under NSF 23-601.

This work is supported by the U.S. Department of Energy, Office of Science, Office of High-Energy Physics, under Contract No. DE–AC02–05CH11231, and by the National Energy Research Scientific Computing Center. Additional support for DESI was provided by the U.S. National Science Foundation, Division of Astronomical Sciences under Contract No. AST-0950945; the Science and Technology Facilities Council of the United Kingdom; the Gordon and Betty Moore Foundation; the Heising-Simons Foundation; the French Alternative Energies and Atomic Energy Commission; the National Council of Humanities, Science and Technology of Mexico; the Ministry of Science, Innovation and Universities of Spain, and by the DESI Member Institutions. 

The authors are honored to conduct research on I'oligam Du'ag, a mountain particularly significant to the Tohono O’odham Nation.

For more information, visit desi.lbl.gov.

\bibliographystyle{aasjournalv7}
\bibliography{references}

\printaffiliations

\end{document}